\documentclass{article}

\usepackage[final]{ml4mols_2020}

\usepackage[utf8]{inputenc} % allow utf-8 input
\usepackage[T1]{fontenc}    % use 8-bit T1 fonts
\usepackage{hyperref}       % hyperlinks
\usepackage{url}            % simple URL typesetting
\usepackage{booktabs}       % professional-quality tables
\usepackage{amsfonts}       % blackboard math symbols
\usepackage{nicefrac}       % compact symbols for 1/2, etc.
\usepackage{microtype}      % microtypography

\usepackage{amsmath}

\usepackage[dvipsnames]{xcolor}
\usepackage{multirow}
\usepackage{graphicx}
\usepackage{cancel}
\usepackage{wrapfig}

\title{Bayesian GNNs for Molecular Property Prediction}

\author{%
  George Lamb \\
  University College London\\
  \texttt{william.lamb.19@ucl.ac.uk} \\
   \And
   Brooks Paige \\
   University College London \\
   \texttt{b.paige@ucl.ac.uk} \\
}

\begin{document}

\maketitle

\begin{abstract}
GNNs for molecular property prediction are frequently underspecified by data and fail to generalise to new scaffolds at test time. A potential solution is Bayesian learning, %in which we consider many possible hypotheses and represent model uncertainty. 
which can capture our uncertainty in the model parameters.
This study benchmarks a set of Bayesian methods applied to a directed MPNN, using the QM9 regression dataset. We find that capturing uncertainty in both readout and message passing parameters yields enhanced predictive accuracy, calibration, and performance on a downstream molecular search task.
\end{abstract}

\section{Introduction}
\label{intro}

Graph neural networks (GNNs) are the state-of-the-art approach to molecular property prediction \citep{duvenaud2015convolutional, gilmer2017neural, wu2018moleculenet, yang2019analyzing}. A GNN operates on the graph structure of a molecule in two phases. In the message passing phase, a molecular representation is learned by passing messages between atom or bond states. In the readout phase, a feed forward network (FFN) converts this representation into a prediction.

\textbf{Motivation}. The particular challenges of molecular property prediction marry well with the potential advantages of Bayesian learning. Generalisation is made difficult in cheminformatics by the concept of a molecular scaffold: the structural core of a compound to which functional groups are attached. Highly parameterised GNNs are prone to over-fit to training scaffolds, learning a poor molecular representation and failing to generalise at test time \citep{yang2019analyzing}. Models are at risk of returning over-confident predictions when operating on new scaffolds, conveying little of the uncertainty associated with a new chemical space. Poorly quantified uncertainty makes it especially challenging to evaluate
model robustness and out-of-domain applicability \citep{hirschfeld2020uncertainty}. We believe the best answer to these deficiencies is Bayesian modelling. Whereas a `classical' neural network bets everything on one hypothesis, a Bayesian approach builds a predictive distribution by considering every possible setting of parameters. Bayesian marginalisation can improve the calibration \citep{maddox2019simple} and accuracy \citep{izmailov2019subspace} of deep neural networks underspecified by data.

\textbf{Related work}. Two recent studies are particularly pertinent. Firstly, \cite{hirschfeld2020uncertainty} benchmark a set of methods for uncertainty quantification in molecular property prediction using the same GNN architecture that we employ in this paper. They find no consistently leading method, though replacing readout with a Gaussian process (GP) or random forest leads to reasonable performance across evaluation metrics. We extend the work of \citeauthor{hirschfeld2020uncertainty} by considering four additional Bayesian methods (SWAG, SGLD, BBP and DUN). Secondly, \cite{hwang2020benchmark} benchmark a set of Bayesian GNNs for molecular property prediction, assessing calibration and predictive accuracy across four classification datasets. They find that Stochastic Weight Averaging (SWA) and SWA-Gaussian (SWAG) demonstrate superior performance across metrics and datasets. We extend the work of \citeauthor{hwang2020benchmark} by (i) working in the regression setting where aleatoric and epistemic uncertainty are more explicitly separable, (ii) directly comparing a Bayesian readout phase with a full Bayesian GNN, and (iii) featuring a downstream molecular search experiment.

We release PyTorch code at \url{https://github.com/georgelamb19/chempropBayes}.
\begin{table}
\caption{Accuracy (measured by Mean Rank), Miscalibration Area (MA) and Search Scores. For single models we present the mean and standard deviation across 5 runs. MAs are computed with post-hoc $t$-distribution likelihoods and presented $\times 10^{2}$. Search Scores equate to the \% of the top 1\% of molecules discovered after 30 batch additions. All Search Scores are computed for single models.}
\centering
\label{table:main}
\begin{tabular}{l c c c c c c}
\toprule
\multirow{2}{*}{Method} & \multicolumn{2}{c}{Accuracy (Mean Rank)} & \multicolumn{2}{c}{Miscalibration Area} & \multicolumn{2}{c}{Search Score} \\ 
\cmidrule{2-7} & Single model & Ensmbl. & Single model & Ensmbl. & Greedy & Thompson \\ 
\midrule
MAP     & 4.08 $\pm$ 0.16 & 4.00  & \textcolor{white}{0}4.20 $\pm$ 0.42 &   \textbf{13.97} & 72.22 $\pm$ 0.57 & n/a \\

GP      & 3.87 $\pm$ 0.42 & 3.17  & \textcolor{white}{0}9.12 $\pm$ 0.98 &   20.28 & 75.22 $\pm$ 1.31 & 75.86 $\pm$ 0.85  \\

DropR   & 7.05 $\pm$ 0.15 & 7.00  & 14.59 $\pm$ 0.37 &  22.09 & 75.38 $\pm$ 1.32 & 76.02 $\pm$ 1.09  \\

DropA   & 7.87 $\pm$ 0.19 & 8.00  & 16.58 $\pm$ 0.47 &   21.98 & \textbf{77.52} $\pm$ 0.77 & \textbf{77.34} $\pm$ 0.88 \\

SWAG    & 3.55 $\pm$ 0.12 & 3.25  & \textcolor{white}{0}9.29 $\pm$ 1.78 &   17.89 & 73.48 $\pm$ 0.75 & 73.14 $\pm$ 0.59  \\

SGLD    & 3.23 $\pm$ 0.51 & 3.75  & \textcolor{white}{0}\textbf{1.79} $\pm$ 1.03 &   14.63 & 69.12 $\pm$ 1.13 & 69.70 $\pm$ 1.31  \\

BBP     & \textbf{1.95} $\pm$ 0.40 & \textbf{1.75}  & \textcolor{white}{0}4.22 $\pm$ 0.57 &   16.51 & 62.27 $\pm$ 6.42 & 61.63 $\pm$ 3.87 \\

DUN     & 4.40 $\pm$ 0.25 & 5.08  & \textcolor{white}{0}4.36 $\pm$ 0.39 &   16.18 & - & -  \\

\bottomrule
\end{tabular}
\end{table}

\section{Model and Data}
\label{model}

\textbf{The D-MPNN}. Our GNN is a directed message passing neural network (D-MPNN) \citep{yang2019analyzing}, a variant of the MPNN family \citep{gilmer2017neural}. The D-MPNN consistently matches or outperforms previous GNN architectures across datasets and splits types \citep{yang2019analyzing}. It has also demonstrated promise in a proof-of-concept antibiotic discovery pipeline \citep{stokes2020deep}. The D-MPNN is the core of the Chemprop project (\url{https://chemprop.readthedocs.io}).

\textbf{QM9}. We perform all experiments on QM9. QM9 contains 12 geometric, energetic, electronic and thermodynamic properties for 133,885 small molecules \citep{ramakrishnan2014quantum}. 
Assessments of uncertainty calibration in Bayesian deep learning tend to focus on classification tasks \citep{lakshminarayanan2017simple, maddox2019simple}. We complement previous studies by exploring calibration and uncertainty quantification in a real-valued regression setting.

\section{Methods}
\label{methods}

We implement eight separate methods. \textbf{MAP}: Our baseline is classical \textit{maximum a posteriori} training, in which we find the regularised maximum likelihood solution. \textbf{GP}: We replace the final layer of the readout FFN with a variational GP and train the resulting model end-to-end (deep kernel learning). The GP is a non-parametric Bayesian method which captures uncertainty in functional form. \textbf{DropR}: MC dropout uses a spike and slab variational distribution to view test time dropout as approximate variational inference \citep{gal2016dropout}. `DropR' is the implementation of MC dropout across readout FFN layers. \textbf{DropA}: We separately implement MC dropout over the full GNN. \textbf{SWAG}: Stochastic Weight Averaging (SWA) \citep{izmailov2018averaging} computes an average of SGD iterates with a high constant learning rate schedule, providing improved generalisation. We implement SWA-Gaussian \citep{maddox2019simple}, which builds on SWA by computing a `low rank plus diagonal' covariance. \textbf{SGLD}: Stochastic Gradient Langevin Dynamics \citep{welling2011bayesian} uses first order Langevin dynamics in the stochastic gradient setting. SGLD is a Markov Chain Monte Carlo (MCMC) method. Within this class of methods Hamiltonian Monte Carlo (HMC) \citep{Neal94} is the gold standard, but requires full gradients which are intractable for modern neural networks. \cite{chen2014stochastic} propose Stochastic Gradient HMC (SGHMC), but in practice tuning this method can be challenging. \textbf{BBP}: Variational Bayesian (VB) methods fit a variational approximation to the true posterior by minimising a Kullback–Leibler (KL) divergence or equivalently maximising an evidence lower bound (ELBO). Bayes by Backprop (BBP) \citep{blundell2015weight} assumes a fully factorised Gaussian posterior and utilises a reparameterisation trick to sample gradients; we also use `local reparameterisation' \citep{kingma2015variational} as a variance reduction technique.
% An alternative VB algorithm would be Stein Variational Gradient Descent \citep{liu2016stein}, which iteratively transports a set of particles to match the target distribution. 
\textbf{DUN}: As an addition to the set of established methods above we implement a novel depth uncertainty network (DUN), which permits inference over both weights and the number of message passing iterations. Our DUN combines Bayes by Backprop with the `vanilla' DUN proposed by \cite{antoran2020depth}, and is introduced in Appendix \ref{DUN}. 
For context, Bayesian modelling is reviewed in Appendix \ref{bayesian_modelling}.
\begin{figure}
\centering
\includegraphics[width=0.47\textwidth]{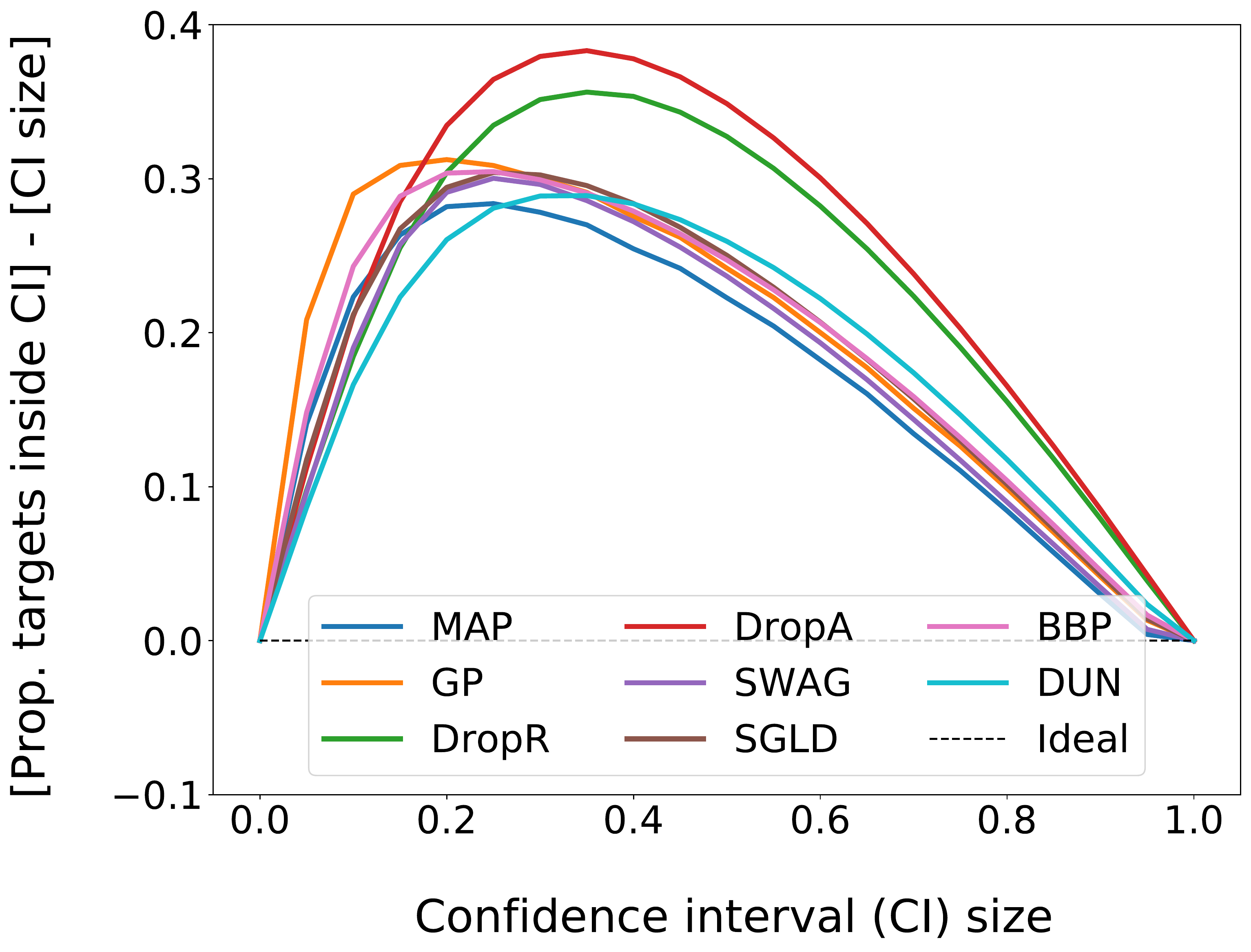}
\hfill
\includegraphics[width=0.47\textwidth]{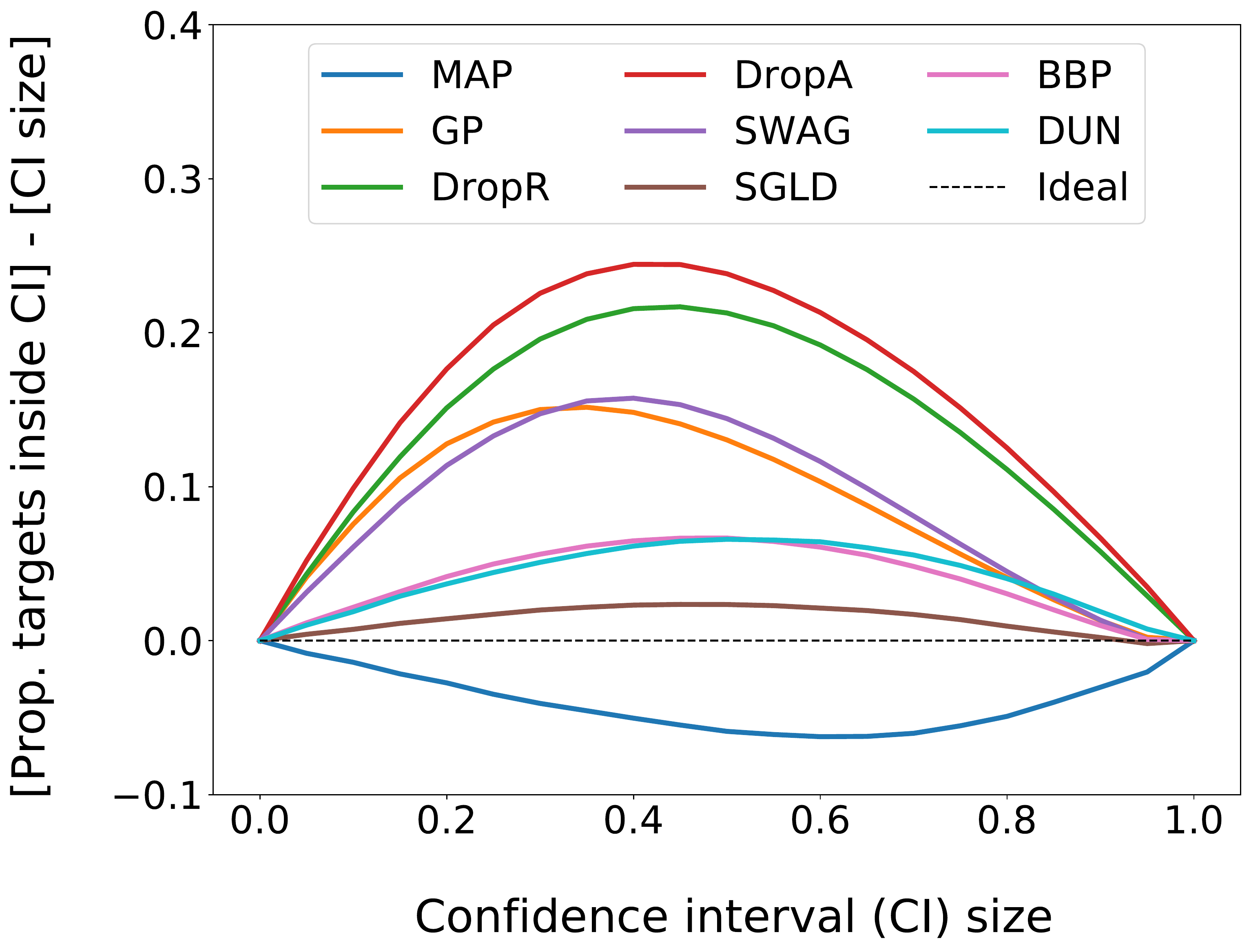}
\caption{Reliability diagrams for single models given Gaussian likelihoods (left) and post-hoc $t$-distribution likelihoods (right). Each line on the diagrams is the average of 5 runs.}
\label{fig:reliability_show}
\end{figure}

\section{Experiments}
\label{experiments}

\subsection{Predictive accuracy and calibration}

\textbf{Framework}. We perform 5 runs for each of the 8 methods, corresponding to different random seeds for weight initialisation. The runs enable an analysis of both `single' models and model ensembles, the latter incorporating multiple posterior basins of attraction. We analyse single models by averaging \textit{scores} across runs, computing a mean and standard deviation. We form a model ensemble by averaging \textit{predictive distributions} across runs, constituting a second layer of Bayesian model averaging. For calibration analysis we model aleatoric noise; a scalar noise per QM9 property is learned within the D-MPNN. Each posterior sample yields an individual Gaussian predictive distribution, representing aleatoric uncertainty. The full Bayesian predictive distribution is a mixture of Gaussians, representing aleatoric and epistemic uncertainty. We create our own data split with [train/val/test] proportions $[0.64,0.16,0.20]$ using Chemprop's `scaffold split' function, which partitions molecules into bins based on their Murcko scaffold. Method implementation is detailed in Appendix \ref{implementation}.

\textbf{Evaluation}. We measure the mean absolute error (MAE) of Bayesian predictive means and rank methods for each of the 12 QM9 tasks. The mean rank across 12 tasks is our chief accuracy evaluation metric. To assess calibration we generalise reliability diagrams \citep{guo2017calibration} to the regression setting and aggregate QM9 tasks. We consider confidence intervals (CIs) around the Bayesian predictive mean. CI size is plotted on the $x$-axis. On the $y$-axis we plot the proportion of molecules in our test set falling within each CI, minus CI size. A perfectly calibrated model is represented by the line $y=0$. We summarise performance on the reliability diagram by computing miscalibration area (MA); the average absolute difference between confidence and accuracy.

\textbf{Results (accuracy)}. Results are presented in Table \ref{table:main} and Appendix \ref{granular}. The leading methods in both single and ensemble settings are BBP, SGLD, SWAG and GP. SGLD and SWAG may suffer slightly versus BBP because they employ vanilla SGD optimisation rather than Adam. SGLD has a higher rank in the single model setting where it is distinguished by its ability to explore multiple posterior modes. Note that the GP captures uncertainty only in readout. Dropout performance is poor, which perhaps could be attributed to an insufficiently large network.
DUN accuracy results should be considered in light of the fact that the variational posterior over depths collapses to $d=5$ (we consider depths of $1$ to $5$), indicating that it has likely failed to capture the true posterior correctly.

\textbf{Results (calibration)}. Reliability diagrams are shown in Figure \ref{fig:reliability_show} and Appendix \ref{reliability_appendix}. With original Gaussian likelihoods we observe pathological underconfidence across methods. We find that this universal underconfidence is driven by overestimated aleatoric uncertainty, a consequence of heavy-tailed residual distributions containing extreme outliers. We improve calibration by fitting post-hoc $t$-distribution likelihoods to training residuals. MA results for post-hoc $t$-distribution likelihoods are shown in Table \ref{table:main}. The post-hoc results motivate re-training with a gamma prior over the precision of our Gaussian likelihood function; placing a prior Gam$(\tau|a,b)$ over $\tau$ and integrating out the precision we obtain a marginal distribution which, after conventional reparameterisation, equates to the $t$-distribution (see \citet[section 2.3.7]{bishop2006pattern}).
We leave this to future work.

\subsection{Molecular search}

\textbf{Framework.} We follow the approximate Bayesian optimisation setup in \cite{hernandez2017parallel}, running both Thompson sampling and greedy trials. Given an unlabelled dataset, the goal is to discover molecules with the largest values of some target property in as few evaluations as possible. At each Thompson iteration we: (i) draw $S$ posterior samples to obtain $S$ deterministic regressors; (ii) for each sample find the molecule with the largest predicted target value, yielding a total batch of $S$ molecules; (iii) query said batch and add it to the labelled training set. Our dataset is a 100k subset of QM9 and our target is the first QM9 property, `mu'. We begin with 5k labelled molecules (selected uniformly at random) and make 30 batch additions with $S=50$. We perform 5 runs per method, corresponding to different base labelled sets and random weight initialisations.

\begin{wrapfigure}{r}{0.47\textwidth}
\begin{center}
\includegraphics[width=0.47\textwidth]{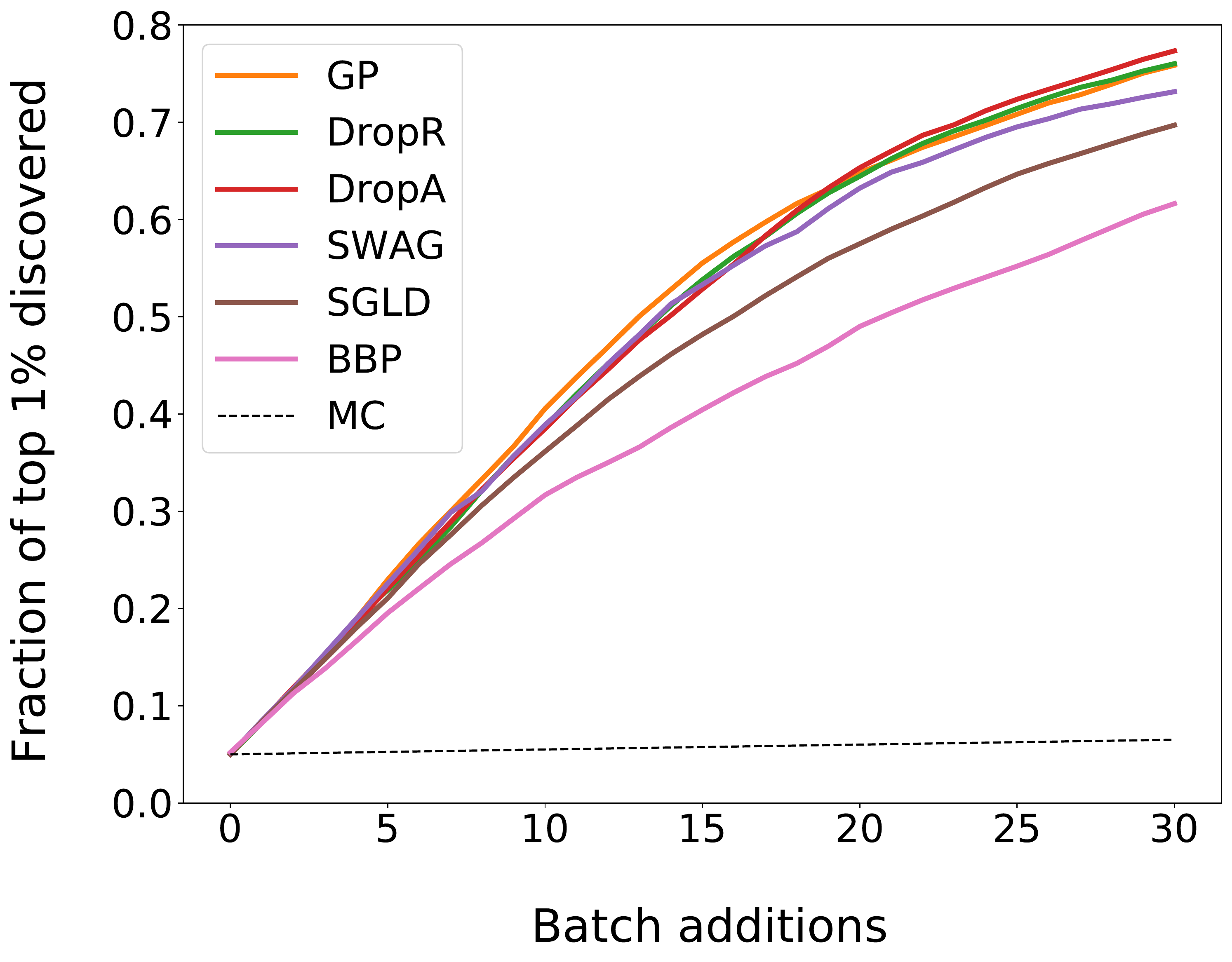}
\end{center}
\caption{Search trajectories for Thompson sampling. Fractions are averaged over 5 runs.}
\label{fig:thompson}
\end{wrapfigure}

\textbf{Evaluation}. After each batch addition we measure the fraction of the top 1\% of molecules discovered. The final metric used to compare methods is the fraction discovered following 30 batch additions, at the close of the experiment. At this point there are 6.5k labelled molecules.

\textbf{Results}. Search scores are presented in Table \ref{table:main}. Thompson sampling trajectories are shown in Figure \ref{fig:thompson} alongside a Monte Carlo baseline. We omit DUN given the collapse of the posterior over depths. As explained in \citet{hernandez2017parallel}, Thompson sampling uses epistemic variance in the Bayesian predictive distribution to perform exploration. In contrast, greedy search selects molecules using the Bayesian predictive mean and exercises pure exploitation. Across methods we find no notable difference between Thompson and greedy search scores. This likely reflects reduced epistemic uncertainty; having randomly selected a subset of 5k molecules to initially label we are operating `in-distribution'. We considered smaller initially labelled sets but found BBP performed particularly poorly. Tuning BBP, SGLD and SWAG without a large validation set is challenging. In contrast, dropout methods and the GP demonstrate robustness to dataset size and hyperparameter settings. The particular success of dropout might also be attributed to its regularising effect.
\section{Discussion and Future Work}
\label{discussion}

The most performant methods involve Bayesian message passing as well as a Bayesian FFN. We conclude that there is meaningful and useful epistemic uncertainty to be captured in learned molecular representations as well as in readout.

When applied to the full QM9 dataset, BBP, SGLD and SWAG enhance accuracy versus a MAP baseline. However, in the context of molecular search the sensitivity of these methods is limiting. Our recommendations follow the observed trends. For precise property prediction with $>10,000$ labelled molecules we suggest experimenting with BBP, SGLD, and SWAG. For molecular search, the robustness of dropout and deep kernel learning to different dataset sizes and hyperparameter settings is advantageous. Our results suggest single model SGLD is the best method for obtaining calibrated uncertainty estimates, though this is likely to be a task-specific phenomenon; extreme outlying residuals are still affecting calibration results despite post-hoc $t$-distribution likelihoods.

We identify three avenues for future work: (i) benchmarking Bayesian GNNs on the complete MoleculeNet dataset collection (see \href{http://moleculenet.ai/datasets-1}{\underline{here}}); the majority of these datasets contain $<10,000$ molecules; (ii) adapting our D-MPNN by placing a gamma prior over the Gaussian likelihood precision, increasing network size for dropout, and experimenting with larger depths (following the DUN posterior collapse we trial depths up to $d=8$ and find accuracy increases monotonically); and (iii) incorporating meta-learning into Bayesian GNNs to improve initialisation in search tasks; meta-initialisations enable rapid learning in low resource settings \citep{nguyen2020meta}.

\newpage
\begin{ack}

WL thanks Aneesh Pappu, Kush Madlani and Udeepa Meepegama for stalwart support, stimulating discussion and occasional commiseration. WL thanks Rob Lewis for blazing a trail from consulting to computer science. BP is supported by The Alan Turing Institute under the UK Engineering and Physical Sciences Research Council (EPSRC) grant no. EP/N510129/1.

\end{ack}

%%%%%%%%%%%%%%%%%%%%%%%%%% references section

\small

\bibliographystyle{apalike}
\bibliography{references}

%%%%%%%%%%%%%%%%%%%%%%%%%% appendix

\newpage
\appendix
\normalsize
\section*{Appendices}

The appendices are structured as follows:
\begin{itemize}
    \item Appendix \ref{bayesian_modelling} reviews the main concepts underlying Bayesian modelling, for readers less familiar with the Bayesian framework.
    \item Appendix \ref{DUN} introduces our depth uncertainty network (DUN) which permits inference over both model weights and the number of message passing iterations.
    \item Appendix \ref{implementation} describes the implementation of methods, and explains key hyperparameter choices.
    \item Appendix \ref{granular} contains granular predictive accuracy results (scaled MAE by task).
    \item Appendix \ref{reliability_appendix} contains a full set of reliability diagrams. Three diagram pairs correspond to (i) Gaussian likelihoods, (ii) post-hoc $t$-distribution likelihoods, and (iii) omission of modelled aleatoric noise.
\end{itemize}

\section{Bayesian Modelling}
\label{bayesian_modelling}

\cite{wilson2020case} emphasises that the distinguishing property of a Bayesian approach is marginalisation rather than optimisation. A Bayesian approach forms a predictive distribution by marginalising over different parameter settings, each weighted by their posterior probability. In contrast, classical learning involves maximising a posterior.

\subsection{Bayesian marginalisation}

We consider the case of Bayesian parametric regression. Given inputs $X$ and outputs $Y$, we desire the parameters $\omega$ of a function $f^{\omega}(\cdot)$ that is likely to have generated our inputs. We place a prior $p(\omega)$ over the space of possible parameter settings, representing our \textit{a priori} belief about which parameters are likely to have generated the data. To transform the prior distribution in light of the observed data we define a likelihood distribution $p(y|x,\omega)$, the probabilistic model by which inputs generate outputs for parameter settings $\omega$. We look for the posterior distribution over the space of parameters by invoking Bayes' theorem:
\begin{equation*}
    p(\omega|\mathcal{D}) = \frac{p(Y|X,\omega)p(\omega)}{p(Y|X)}.
\end{equation*}

A predictive distribution is obtained by marginalising over $\omega$:
\begin{equation}\label{eq:predictive}
    p(y|x,\mathcal{D}) = \int p(y|x,\omega)p(\omega|\mathcal{D}) d \omega.
\end{equation}
Equation (\ref{eq:predictive}) is a Bayesian model average (BMA), representing model uncertainty.

\subsection{Bayesian deep learning}

Modern neural networks often contain millions of parameters. The posterior over these parameters is generally intractable. In Bayesian deep learning we deal with the problem of inference by making two, layered approximations. Firstly, we approximate the Bayesian posterior. Methods differ with respect to posterior approximation. Secondly, we approximate the Bayesian integral (\ref{eq:predictive}) by Monte Carlo (MC) sampling. MC integration is common across methods.

With $q(\omega|\mathcal{D})$ our approximate posterior, the MC BMA is:
\begin{equation*}
    p(y|x,\mathcal{D}) \approx \frac{1}{J} \sum_{j=1}^{J}
    p(y|x,\omega_{j}), \hspace{3mm}
    \omega_{j} \sim q(\omega|\mathcal{D}).
\end{equation*}
Following MC integration, we have approximated the true posterior with a set of point masses, where their locations are given by samples from $q(\omega|\mathcal{D})$:
\begin{equation*}
    p(\omega|\mathcal{D}) \approx \frac{1}{J}
    \sum_{j=1}^{J} \delta (\omega = \omega_{j}), \hspace{3mm}
    \omega_{j} \sim q(\omega|\mathcal{D}).
\end{equation*}
\section{The Depth Uncertainty Network}
\label{DUN}

Here we consider capturing uncertainty in model weights \textit{and} the number of message passing iterations. For simplicity we refer to the latter parameter as `depth'. There is motivation to acknowledge and capture uncertainty in the MPNN depth parameter. Different depths allow hidden states to represent different sized molecular substructures. Incorporating different sized spheres of representation at test time may enhance predictive accuracy.

\subsection{Depth uncertainty in an FFN}

\cite{antoran2020depth} perform inference over the depth of an FFN. Different depths correspond to subnetworks which share weights. Exploiting the sequential structure of FFNs, \citeauthor{antoran2020depth} evaluate a training objective and make predictions with a single forward pass.

\citeauthor{antoran2020depth} define a categorical prior over network depth, $p_{\beta}(d)=\text{Cat}(d|\{ \beta_{i} \}_{i=0}^{D})$. They parameterise the likelihood for each depth using the corresponding subnetwork's output: $p(\mathbf{y}|\mathbf{x},d=i,\omega)=p(\mathbf{y}|f_{D+1}(\mathbf{a}_{i},\omega))$. Here, $f_{D+1}(\cdot)$ is an output layer and $\mathbf{a}_{i}$ the activation at depth $i \in [0,D]$ given input $\mathbf{x}$ and weight configuration $\omega$. For a given weight configuration, the likelihood for each depth and consequently the model's marginal log likelihood (MLL) can be computed from a single forward pass. The MLL is computed as
\begin{equation*}
    \log p(\mathcal{D},\omega) = \log \sum_{i=0}^{D} \Bigg( p_{\beta}(d=i) \cdot 
    \prod_{n=1}^{N} p(\mathbf{y}^{(n)}|\mathbf{x}^{(n)},d=i,\omega) 
    \Bigg).
\end{equation*}
The posterior over depth is a tractable categorical distribution which tells us how well each subnetwork explains the data given some set of weights $\omega$:
\begin{equation*}
    p(d|\mathcal{D},\omega) = \frac{p(\mathcal{D}|d,\omega) p_{\beta}(d)}{p(\mathcal{D},\omega)}.
\end{equation*}

\citeauthor{antoran2020depth} try learning weights by maximising the MLL directly using backpropagation and the \textit{log-sum-exp} trick, but find the posterior collapses to a delta function over an arbitrary depth. This is explained by the gradients of each subnetwork being weighted by that subnetwork's posterior mass, leading to local optima where all but one subnetwork's gradients vanish.

The solution is to separate the optimisation of network weights from the posterior distribution as in the expectation maximisation (EM) algorithm for latent variable models. \citeauthor{antoran2020depth} achieve this by performing stochastic gradient variational inference. They introduce a variational posterior $q_{\alpha}(d)=\text{Cat}(d|\{ \alpha_{i} \}_{i=0}^{D})$. They learn variational parameters $\alpha$ and weights $\omega$ by maximising the following ELBO:
\begin{equation*}\label{deep_elbo1}
    \log p(\mathcal{D},\omega) \geq \mathcal{L}(\alpha, \omega) = \sum_{n=1}^{N}
    \mathbb{E}_{q_{\alpha}(d)} \big[ \log p(\mathbf{y}^{(n)}|\mathbf{x}^{(n)},d,\omega) \big]
    -
    \text{KL}(q_{\alpha}(d)\hspace{1mm}||\hspace{1mm}p_{\beta}(d)).
\end{equation*}
Maximising this ELBO by taking gradients actually constitutes exact inference. The ELBO is convex w.r.t. $\alpha$ because the variational and true posteriors are categorical. The variational family contains the exact posterior, thus at the maxima $q_{\alpha}(d)=p(d|\mathcal{D},\omega)$.

\subsection{Incorporating uncertainty in model weights}

Our depth uncertainty network (DUN) combines the model above with Bayes by Backprop. We assume a fully factorised variational posterior over depth and neural network weights. We derive an ELBO by expanding the KL divergence:
\begin{equation*}
\begin{split}
    \text{KL} \big[ &q(d|\alpha) q(\omega|\theta) \hspace{1mm}||\hspace{1mm} p(d|\mathcal{D},\omega) p(\omega|\mathcal{D}) \big]\\
    &=
    \text{KL} \big[ q(d|\alpha) \hspace{1mm}||\hspace{1mm} p(d) \big] +
    \text{KL} \big[ q(\omega|\theta) \hspace{1mm}||\hspace{1mm} p(\omega) \big] -
    \mathbb{E}_{q(d|\alpha)q(\omega|\theta)}[\log p(\mathcal{D}|d,\omega)] +
    \log p(\mathcal{D})\\
    &=
    -\mathcal{L}(\alpha, \theta) + \log p(\mathcal{D}).
\end{split}
\end{equation*}
$\mathcal{L}(\alpha, \theta)$ here is the ELBO. Due to the non-negativity of the KL divergence, $\mathcal{L}(\alpha, \theta) \leq \log p(\mathcal{D})$.

We can learn variational parameters $\alpha$ and $\theta$ by maximising the ELBO using backpropagation. The KL term involving categorical distributions and the expectation of the log likelihood w.r.t. the posterior over depth can be computed analytically with a single forward pass. Expectations w.r.t. the variational posterior $q(\omega | \theta)$ are estimated as in Bayes by Backprop, by sampling unbiased gradients.

In practice we can use mini-batches, estimating the ELBO as follows:
\begin{equation}\label{eq:dun_loss}
\begin{split}
    \mathcal{L}(\omega, \alpha, \theta)
    \approx
    \frac{N}{B} \sum_{n=1}^{B} \sum_{i=0}^{D}
    \big(
    \log p(\mathbf{y}^{(n)}|&\mathbf{x}^{(n)}, d = i, \omega) \cdot \alpha_{i}
    \big)\\
    &- \log
    \frac{q(\omega | \theta)}{p(\omega)}
    -
    \sum_{i=0}^{D}
    \big( \alpha_{i} \log \frac{\alpha_{i}}{\beta_{i}}
    \big).
\end{split}
\end{equation}

At test time, predictions for new data are made by the following Bayesian model average:
\begin{equation*}\label{eq:dun_bma}
\begin{split}
    p(\mathbf{y}^{*}|\mathbf{x}^{*},\mathcal{D}) &\approx \frac{1}{J} \sum_{j=1}^{J}
    \sum_{i=0}^{D}
    p(\mathbf{y}^{*}|\mathbf{x}^{*},d=i,\omega_{j}) q_{\alpha}(d=i),\\
    \omega_{j} &\sim q(\omega|\theta).
\end{split}
\end{equation*}
\section{Implementation of Methods}
\label{implementation}

In this section we describe the implementation of methods for the predictive accuracy and calibration experiment. Unless otherwise specified, hyperparameters are set to optimise validation MAE (averaged across QM9 tasks) following grid-search. An exhaustive list of the hyperparameter settings for all our experiments can be found in the file \verb+chempropBayes/scripts/bayesHyp.py+.

\subsection{MAP}

In order to learn aleatoric noise we instantiate a log standard deviation parameter within the D-MPNN model (the log ensures non-negativity). This parameter is a set of 12 scalars, one for each of the QM9 tasks. We henceforth refer to the parameter as `log noise'.

Our full loss object is the negative log likelihood plus the negative log prior. A function to compute the former takes as input predictions, targets and log noise. We place a zero-mean Gaussian prior over each D-MPNN weight and control $\sigma_{\text{prior}}$ via a weight decay hyperparameter $\lambda$ inside our optimiser. In practice we scale the negative log likelihood to a manageable order of magnitude by dividing by the batch size. This scales the relationship between weight decay and our prior sigma. Precisely, we have $\lambda = 1 / \sigma_{\text{prior}}^{2}N$ where $N$ is the training set size.

The default batch size in Chemprop is 50 and we find this works well; we use this batch size across all methods. Our optimiser is Adam. Following grid search we set the weight decay to $\lambda = 0.01$.

Chemprop utilises a `noam' learning rate scheduler with piecewise linear increase and exponential decay (based on the scheduler in \citet{vaswani2017attention}). We train for 200 epochs, using the `noam' scheduler for the first 100. We linearly increase the learning rate from $lr_{\text{min}}$ to $lr_{\text{max}}$ over 2 epochs and decay back to $lr_{\text{min}}$ over the following 98, from which point we remain at $lr_{\text{min}}$. Following grid search we set $lr_{\text{min}}$ = 1e-4 and $lr_{\text{max}}$ = 1e-3. The saved MAP model following each training run is that which achieves the best validation accuracy. We also apply this selection procedure to GP, DropR, DropA, BBP and DUN.

\textbf{Architecture}. We grid search over 24 architectures, exploring combinations of hidden size $h \in \{300,500\}$, message passing depth $d \in \{ 2, 3, 4, 5 \}$ and number of FFN readout layers $L \in \{ 2, 3, 4 \}$. We find that $(h,d,L) = (500,5,3)$ achieves optimal accuracy. We choose not to investigate larger hidden sizes or depths to manage compute requirements. For context, the Chemprop defaults are $(h,d,L) = (300,3,2)$. We maintain a fixed architecture across all methods.

\textbf{Standardising features and targets}. Before training the D-MPNN we standardise atom and bond features in the training set, and apply the same transformation to validation and test molecules. We also standardise training targets, later applying the reverse transform when making predictions on validation or test molecules. Both these standardisations occur across all methods.

\subsection{GP}

Each GP run is initialised with the output of the corresponding MAP run (e.g. GP run 1 is initialised with the output of MAP run 1). We take the pre-trained MAP D-MPNN and replace the final layer of readout with 12 batched stochastic variational GPs (SVGPs), one per task. We train the resulting architecture end-to-end. This end-to-end training is known as deep kernel learning (DKL).

We implement the GPs in GPyTorch, following the example SVGP and DKL implementations as a guide (\url{https://docs.gpytorch.ai}). Our variational distribution is a multivariate normal (`CholeskyVariationalDistribution') with batch shape 12. We use a multitask variational strategy with 1200 inducing points based on methodology in \citet{hensman2013gaussian}. The variational strategy tells us how to transform a distribution $q(u)$ over the inducing point values to a distribution $q(f)$ over the latent function values for some input $x$. 1200 is the maximum feasible number of inducing points given compute constraints (corresponding to 10-15 minutes per epoch on a single GPU node). We note that the closeness of the variational GP approximation to the true posterior increases only with the log of the number of inducing points \citep{matthews2016sparse}.

Each GP is defined by a constant mean and RBF kernel. We train GP hyperparameters, a scalar aleatoric noise per task and D-MPNN weights by minimising a negative variational ELBO. We train with a batch size of 50 for 200 epochs, following the same learning rate profile as for MAP. We use the Adam optimiser. For fair comparison with other methods we regularise D-MPNN weights with a weight decay of $0.01$.

\subsection{DropR, DropA}

The dropout models follow a similar training procedure to MAP; here we highlight differences. For DropR we activate dropout layers following the D-MPNN atom representation step, and following every FFN layer except for the output layer. For DropA we additionally activate dropout layers following D-MPNN hidden state updates.

For both DropR and DropA we grid search over $p \in \{0.1, 0.2, 0.3, 0.4, 0.5 \}$ (these are \textit{dropout} probabilities). For both DropR and DropA the optimal probability is $0.1$. We run the final models with this dropout probability during training and testing.

Both dropout methods take significantly longer to converge than MAP. This is expected given the noise inherent in dropout. We train for 300 epochs, extending the MAP learning rate profile by 100 additional epochs at a fixed learning rate of $lr_{\text{min}}$ = 1e-4. At test time we draw 30 samples.

\subsection{SWAG}

The SWAG implementation is based on code attached to the original SWAG paper \citep{maddox2019simple}. We first build a wrapper around the D-MPNN, referring to the latter as our `base' model. The wrapper contains a list of the parameter objects in the base model (where a parameter `object' is, for example, a weight matrix). Within the wrapper list (which includes log noise) we register buffers for each parameter object to store first and second uncentred moments. During training we `collect' models by looping through parameter objects in the base model and updating buffers in the wrapper list. At test time we generate new sample parameters \textit{directly within the wrapper list}. Because the parameter objects in the list point to the same place in memory as the parameters in the base model, base model parameters also change when we sample.

The starting point for SWAG training is the pre-trained MAP model. We run SWAG training for 100 epochs, collecting one model per epoch after 20 warm-up epochs (thus collecting 80 models in total). We limit the rank of our estimated covariance matrix by using only the last $K=20$ models to compose a deviation matrix (the same setting as in the original paper). For fair comparison with other methods we set weight decay to $\lambda=0.01$.

The performance of the SWAG method is sensitive to learning rates. To prevent spikes in loss during training we make three changes versus MAP. Firstly, we lower the main learning rate. In practice 2e-5 is the highest rate with which we can achieve reasonable validation accuracy during training (SWAG should be run with a constant `high' learning rate). Secondly, we reduce the learning rate even further for log noise; at all times it is one fifth of the learning rate applied to other model parameters. Thirdly, at the start of SWAG training we gradually increase learning rates from 1e-10 up to their maxima over 5 epochs, using a cosine scheduler (the scheduler is not particularly important; we use a cosine scheduler for alignment with SGLD). SWAG's sensitivity is a result of using the SGD optimiser as opposed to Adam (which enjoys momentum and adaptive learning rates). We try SWAG with momentum $\rho \in \{ 0.5, 0.9, 0.99 \}$ but see more volatile loss profiles so momentum is kept at zero. At test time we draw 30 samples.

\subsection{SGLD}

SGLD parameter updates are equivalent to MAP updates with the addition of Gaussian noise. As with MAP, we scale the SGLD loss to a manageable order of magnitude by dividing by the training set size, $N$. This division effectively rescales the SGLD learning rate to be $\epsilon / N$. It follows that we should also divide the variance of our added Langevin noise by $N$. Denoting the batch size as $B$, our parameter update equation is:
\begin{equation*}
\begin{split}
    \Delta \omega_{t} &= \frac{\epsilon}{2N}
    \bigg(
    \nabla \log p(\omega_{t})
    + \frac{N}{B}
    \sum_{i=1}^{B} \nabla \log p(\mathbf{y}_{i}|\mathbf{x}_{i}, \omega_{t})
    \bigg)
    +
    \eta_{t},\\
    \eta_{t} &\sim \mathcal{N}(0,\epsilon / N).
\end{split}
\end{equation*}

We implement SGLD as an optimiser which inherits from PyTorch's SGD base class. The SGLD optimiser loops through parameter groups and adds two gradient terms to the already-computed log likelihood gradients. Firstly, it adds the gradient of the log prior; we parameterise this via a weight decay and set the weight decay to $\lambda = 0.01$ for fair comparison with other methods. Secondly, the optimiser adds appropriately scaled Langevin noise.

The starting point for SGLD is the pre-trained MAP model. We run SGLD with a cyclical cosine learning rate schedule, following the proposal of \cite{zhang2019cyclical}. The idea is that larger steps discover new posterior modes during a period of exploration (effectively a burn-in phase), and that smaller steps characterise each mode. We use PyTorch's `OneCycleLR' scheduler and configure a single cycle as follows: we cosine anneal the learning rate from 1e-10 to a maximum learning rate of 1e-4 over 5 epochs, and then cosine anneal from this maximum to a minimum learning rate of 1e-5 over the following 45 epochs. At all times the learning rate for log noise is one fifth of the main learning rate. We save a posterior sample at the end of each 50 epoch cycle. Given this relatively expensive serial sampling procedure, we collect only 20 samples for SGLD.

\subsection{BBP}

Again, we scale the loss to be a per example measure. Given batch size $B$ the loss function is:
\begin{equation*}
    \mathcal{L}(\omega,\theta) = 
    \frac{1}{N}
    \bigg(
    \log q(\omega|\theta)
    - 
    \log p(\omega)
    -
    \frac{N}{B}
    \sum_{i=1}^{B} \nabla 
    \log p(\mathbf{y}_{i}|\mathbf{x}_{i},\omega)
    \bigg).
\end{equation*}
In practice, we average this loss across 5 forward passes before every backward pass to reduce variance.

To implement BBP, we define a Bayesian linear layer class to replace the existing linear layers in the D-MPNN. Within the Bayesian linear layer we implement the `local reparameterisation trick' \citep{kingma2015variational}. This involves calculating the mean and variance of activations in closed form and sampling activations instead of weights. With the local reparameterisation trick the variance of our MC ELBO estimator scales as $1/B$; sampling weights directly it scales $(B-1)/B$ (with $B$ the batch size). Within each Bayesian linear layer we also compute the KL divergence in closed form using a result from \citet[Appendix B]{kingma2013auto}. Each layer returns standard output as well as a KL. We sum the latter across layers to compute a total KL.

We initialise BBP from the MAP solution and train for 100 epochs at a constant learning rate of 1e-4. We set $\sigma_{\text{prior}}=0.05$ which is approximately equivalent to a weight decay of $0.01$ given our scaled loss. Initialising $\rho$ parameters in the correct range is important for reasonable convergence; we initialise uniformly at random between $-5.5$ and $-5$. Each training run we save the BBP model with the best validation accuracy, where validation accuracy is calculated for the mean of the variational posterior. At test time we draw 30 samples.

\subsection{DUN}

The DUN method is implemented on top of BBP. Our loss is the negative of equation (\ref{eq:dun_loss}) (though we compute an exact BBP KL). As with previous methods, we rescale the loss by dividing by the training set size. The variational categorical distribution $q(d|\alpha)$ is learned as a set of parameters within the D-MPNN, where we use logs to ensure non-negativity. In a single forward pass our DUN D-MPNN returns a BBP KL (the sum of KLs computed in closed form within Bayesian linear layers), a KL of categorical distributions (also computed exactly) and predictions corresponding to five different depths. Our categorical prior is the uniform distribution.

Categorical distributions are over $d \in \{1,2,3,4,5 \}$. Note that in Chemprop the depth $d$ counts the number of message passing steps plus the hidden state initialisation step. We do not exceed $d=5$ for fair comparison; improved DUN performance versus other methods may otherwise be caused by the inclusion of a deeper sub-model alone. Recall that when selecting a model architecture we only grid search up to $d=5$.

We train DUN models for 350 epochs. For the first 100, BBP $\rho$ parameters are frozen at zero (thus our variational posterior over weights is a point mass) and we freeze $q(d|\alpha)$ to be a uniform distribution. This first phase of training is designed to minimise the chance of the variational categorical distribution later collapsing to a single depth. For the first 100 epochs we use the `noam' scheduler with MAP learning rates. After 100 epochs we initialise $\rho$ as in the BBP method and unfreeze our variational categorical parameters. From 100 epochs onward we use a constant learning rate of 1e-4. We save the DUN model achieving the best validation accuracy.

At test time we generate two sets of samples. We draw 30 samples from the marginal posterior over weights, each of which predicts by taking an expectation w.r.t. depth; we use these samples to evaluate DUN accuracy. We also draw 100 samples from the joint posterior over weights and depth, using these to assess uncertainty calibration; the larger number of samples is necessary to minimise discretisation error.

\begin{table}[h]
\vspace{1.6cm}
\caption{MAE for MAP and Bayesian \textbf{model ensembles}. The 5 runs in Table \ref{table:acc_single} constitute an ensemble of 5 models. Results are scaled by task such that predicting with the mean of test targets would yield an MAE score of 100.\vspace{3mm}}
\centering
\label{table:acc_ens}
\begin{tabular}{l r r r r r r r r}
\toprule
Property & \hspace{0.25cm}MAP & \hspace{0.5cm}GP & \hspace{0.1cm}DropR & \hspace{0.1cm}DropA & \hspace{0.15cm}SWAG & \hspace{0.2cm}SGLD & \hspace{0.2cm}BBP & \hspace{0.2cm}DUN\\ 
\midrule
mu    & 45.64 & 45.43 & 49.10 & 52.17 & 45.47 & 45.41 & \textbf{45.13} & 46.35  \\

alpha &  6.70 &  6.68 &  8.20 &  9.22 &  6.49 &  6.56 &  \textbf{6.39} &  6.90  \\

HOMO  & 26.69 & 27.24 & 28.59 & 32.42 & 26.37 & \textbf{26.08} & 26.25 & 26.27  \\

LUMO  &  9.52 &  9.85 & 10.42 & 11.68 &  9.40 &  9.28 &  \textbf{9.25} &  9.51  \\

gap   & 13.37 & 13.75 & 14.47 & 16.56 & 13.12 & \textbf{12.97} & 13.04 & 13.17  \\

R2    & 13.52 & 13.51 & 15.50 & 16.55 & 13.45 & 13.41 & \textbf{13.34} & 13.99  \\

ZPVE  &  1.57 &  \textbf{1.38} &  3.27 &  3.49 &  1.56 &  1.61 &  1.59 &  1.70  \\

Cv    &  5.76 &  5.78 &  7.61 &  8.36 &  5.74 &  5.83 &  \textbf{5.70} &  5.99  \\

U0    &  1.57 &  \textbf{1.30} &  2.71 &  2.76 &  1.86 &  1.95 &  1.51 &  1.92  \\

U     &  1.57 &  \textbf{1.30} &  2.71 &  2.76 &  1.81 &  1.95 &  1.51 &  1.93  \\

H     &  1.57 &  \textbf{1.30} &  2.71 &  2.76 &  1.83 &  1.95 &  1.51 &  1.93  \\

G     &  1.57 &  \textbf{1.30} &  2.71 &  2.76 &  1.83 &  1.94 &  1.51 &  1.92  \\

\textit{All} & 10.76 & 10.73 & 12.33 & 13.46 & 10.74 & 10.75 & \textbf{10.56} & 10.96  \\
\midrule
Mean rank    &  4.00 &  3.17 &  7.00 &  8.00 &  3.25 &  3.75 &  \textbf{1.75} &  5.08  \\

\bottomrule
\end{tabular}
\end{table}
\newpage
\section{Granular Predictive Accuracy Results}
\label{granular}

Tables \ref{table:acc_ens} and \ref{table:acc_single} present granular predictive accuracy results. MAEs are scaled by task such that predicting with the mean of test targets would yield an MAE score of 100. The primary metric in these tables is `mean rank', calculated (per run) by averaging the rank of a method across the 12 tasks. Using mean rank ensures we evenly weight the 12 tasks. To use MAE averaged across tasks (the `All' row in the tables) would be to give a higher weighting to more difficult tasks.

\begin{table}[h]
\vspace{1cm}
\caption{(split table). MAE for MAP and Bayesian \textbf{single models}. Means and standard deviations are computed across 5 runs. Results are scaled by task such that predicting with the mean of test targets would yield an MAE score of 100.\vspace{3mm}}
\centering
\label{table:acc_single}
\begin{tabular}{l r r r r r r r r}
\toprule
\multirow{2}{*}{Property} & \multicolumn{2}{c}{\hspace{0.6cm}MAP} & \multicolumn{2}{c}{\hspace{0.6cm}GP} & \multicolumn{2}{c}{\hspace{0.6cm}DropR} & \multicolumn{2}{c}{\hspace{0.6cm}DropA}\\ 
\cmidrule{2-9}
 \hspace{1.0cm} & \hspace{0.6cm}mean & std  & \hspace{0.6cm}mean & std & \hspace{0.6cm}mean & std & \hspace{0.6cm}mean & std\\ 
\midrule
mu    & 48.41 &  0.43 & 48.95 &  0.29 & 50.42 &  0.67 & 52.77 &  0.37  \\

alpha &  7.89 &  0.15 &  8.01 &  0.31 &  9.32 &  0.22 & 10.12 &  0.12  \\

HOMO  & 30.06 &  0.27 & 31.13 &  0.40 & 30.76 &  0.29 & 33.38 &  0.40  \\

LUMO  & 11.08 &  0.15 & 11.44 &  0.04 & 11.86 &  0.36 & 12.34 &  0.21  \\

gap   & 15.35 &  0.11 & 15.87 &  0.12 & 16.07 &  0.25 & 17.24 &  0.38  \\

R2    & 15.44 &  0.17 & 15.74 &  0.28 & 16.86 &  0.27 & 17.53 &  0.20  \\

ZPVE  &  1.93 &  0.06 &  \textbf{1.74} &  0.14 &  4.72 &  0.27 &  4.60 &  0.20  \\

Cv    &  6.94 &  0.04 &  7.05 &  0.21 &  9.26 &  0.29 &  9.50 &  0.17  \\

U0    &  1.88 &  0.09 &  \textbf{1.55} &  0.19 &  3.58 &  0.15 &  3.76 &  0.10  \\

U     &  1.88 &  0.08 &  \textbf{1.55} &  0.19 &  3.58 &  0.15 &  3.76 &  0.10  \\

H     &  1.88 &  0.08 &  \textbf{1.55} &  0.19 &  3.58 &  0.15 &  3.76 &  0.10  \\

G     &  1.88 &  0.09 &  \textbf{1.55} &  0.18 &  3.58 &  0.15 &  3.76 &  0.10  \\

\textit{All} & 12.05 &  0.05 & 12.18 &  0.14 & 13.63 &  0.15 & 14.37 &  0.16  \\
\midrule
Mean rank    &  4.08 &  0.16 &  3.87 &  0.42 &  7.05 &  0.15 &  7.87 &  0.19  \\

\midrule

\\

\\

\midrule

\multirow{2}{*}{Property} & \multicolumn{2}{c}{\hspace{0.6cm}SWAG} & \multicolumn{2}{c}{\hspace{0.6cm}SGLD} & \multicolumn{2}{c}{\hspace{0.6cm}BBP} & \multicolumn{2}{c}{\hspace{0.6cm}DUN}\\ 
\cmidrule{2-9}
  & mean & std & mean & std & mean & std & mean & std \\ 
\midrule
mu    & 48.03 &  0.33 & 47.80 &  0.33 & \textbf{47.72} &  0.47 & 48.24 &  0.29  \\

alpha &  7.58 &  0.03 &  7.49 &  0.10 &  \textbf{7.43} &  0.15 &  7.84 &  0.15  \\

HOMO  & 29.49 &  0.25 & 28.93 &  0.32 & 29.12 &  0.25 & \textbf{28.64} &  0.17  \\

LUMO  & 10.79 &  0.07 & \textbf{10.49} &  0.08 & 10.53 &  0.13 & 10.54 &  0.02  \\

gap   & 14.98 &  0.08 & 14.63 &  0.10 & 14.68 &  0.08 & \textbf{14.59} &  0.22  \\

R2    & 15.28 &  0.11 & 15.06 &  0.18 & \textbf{15.02} &  0.08 & 15.41 &  0.18  \\

ZPVE  &  1.90 &  0.04 &  1.95 &  0.09 &  1.88 &  0.11 &  2.13 &  0.11  \\

Cv    &  6.80 &  0.05 &  6.79 &  0.10 &  \textbf{6.64} &  0.08 &  6.94 &  0.14  \\

U0    &  2.04 &  0.20 &  2.30 &  0.22 &  1.76 &  0.08 &  2.42 &  0.21  \\

U     &  2.01 &  0.18 &  2.30 &  0.22 &  1.76 &  0.08 &  2.42 &  0.22  \\

H     &  2.02 &  0.19 &  2.30 &  0.22 &  1.76 &  0.09 &  2.42 &  0.22  \\

G     &  2.04 &  0.19 &  2.30 &  0.22 &  1.76 &  0.09 &  2.42 &  0.22  \\

\textit{All} & 11.91 &  0.07 & 11.86 &  0.13 & \textbf{11.67} &  0.09 & 12.00 &  0.06  \\
\midrule
Mean rank    &  3.55 &  0.12 &  3.23 &  0.51 &  \textbf{1.95} &  0.40 &  4.40 &  0.25  \\

\bottomrule

\end{tabular}
\end{table}
\newpage
\section{Reliability Diagrams}
\label{reliability_appendix}

\begin{figure}[h]
\centering
\includegraphics[width=0.52\textwidth]{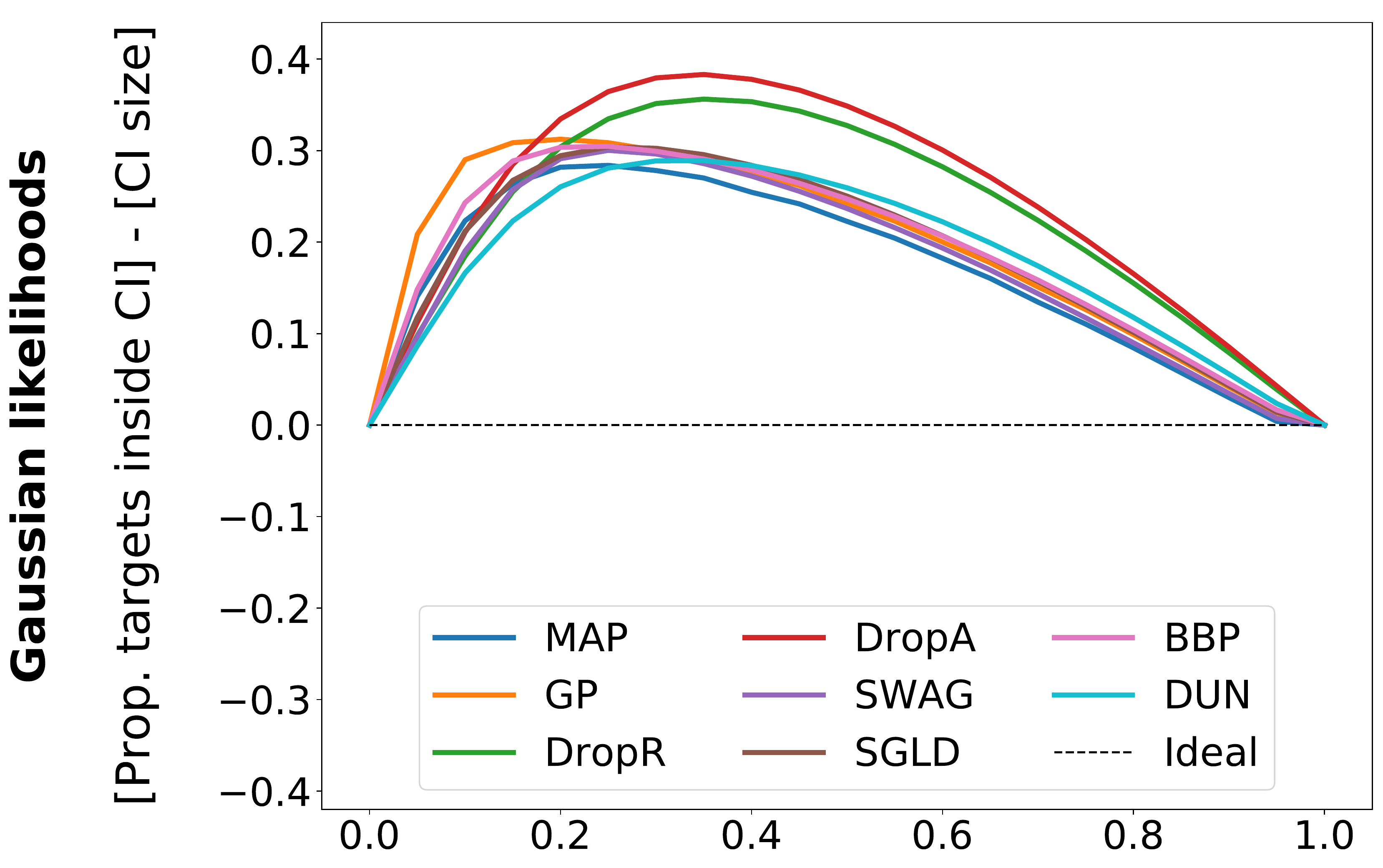}
\hfill
\includegraphics[width=0.4428\textwidth]{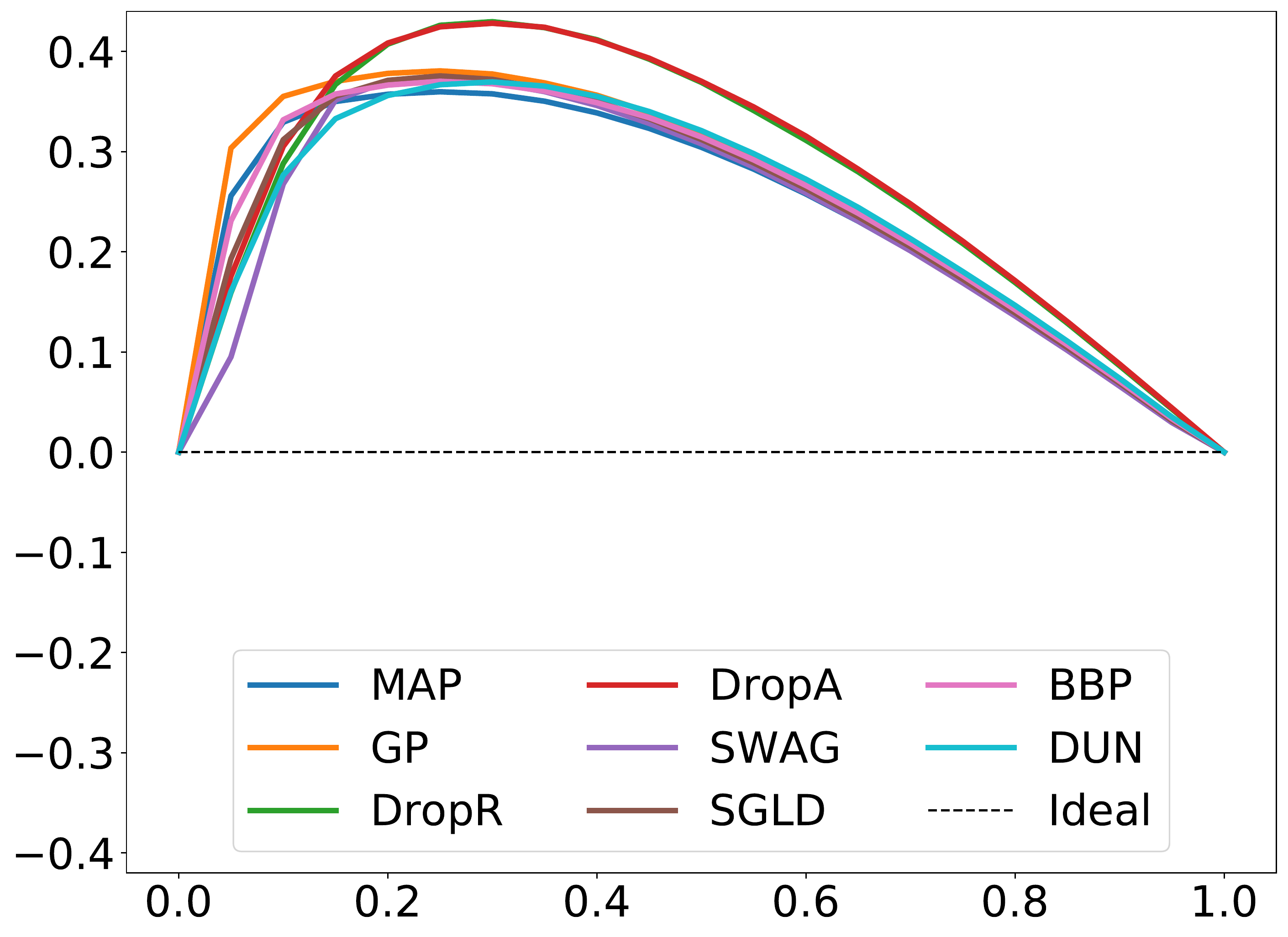}\\
\vspace{3mm}
\includegraphics[width=0.52\textwidth]{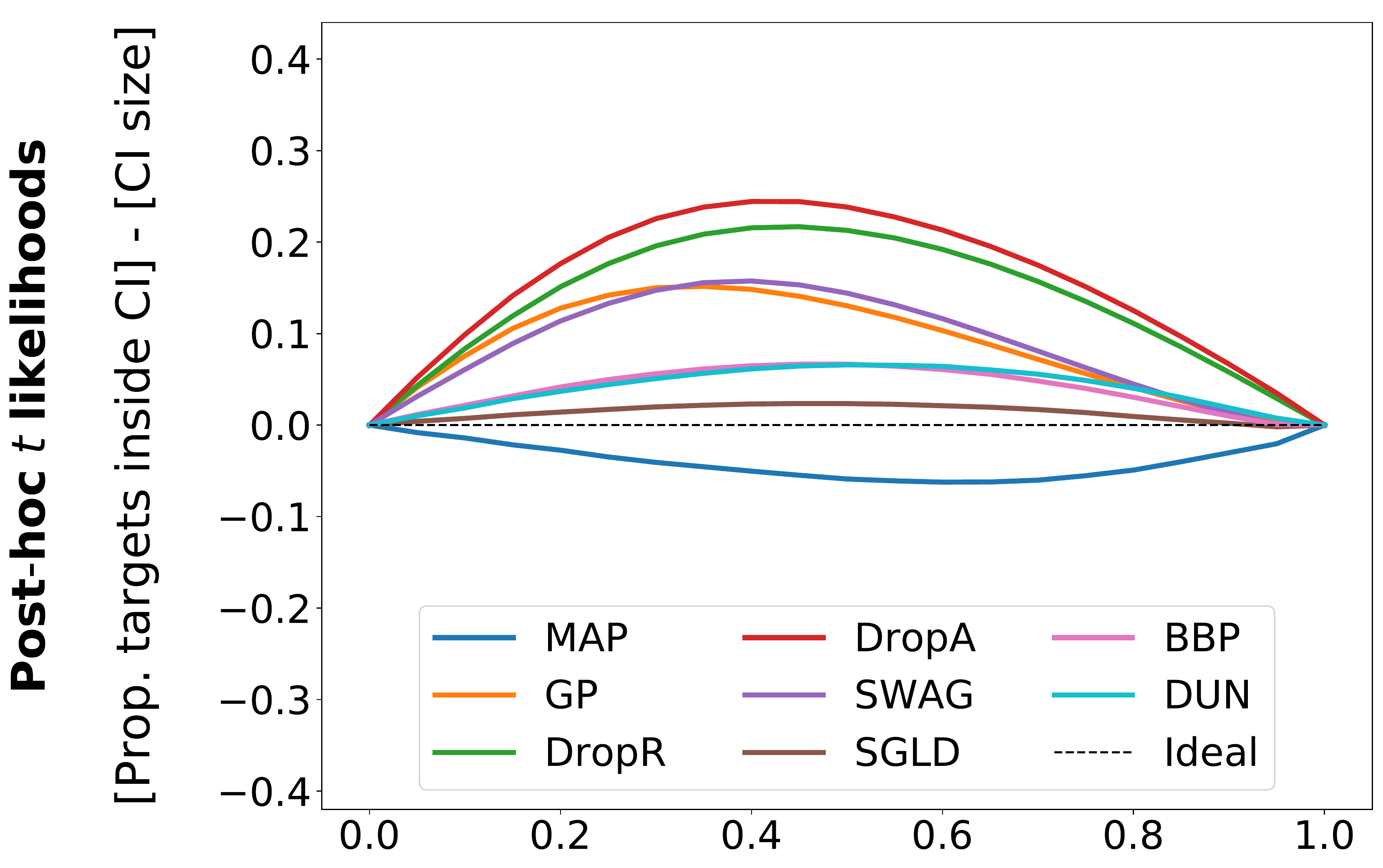}
\hfill
\includegraphics[width=0.4428\textwidth]{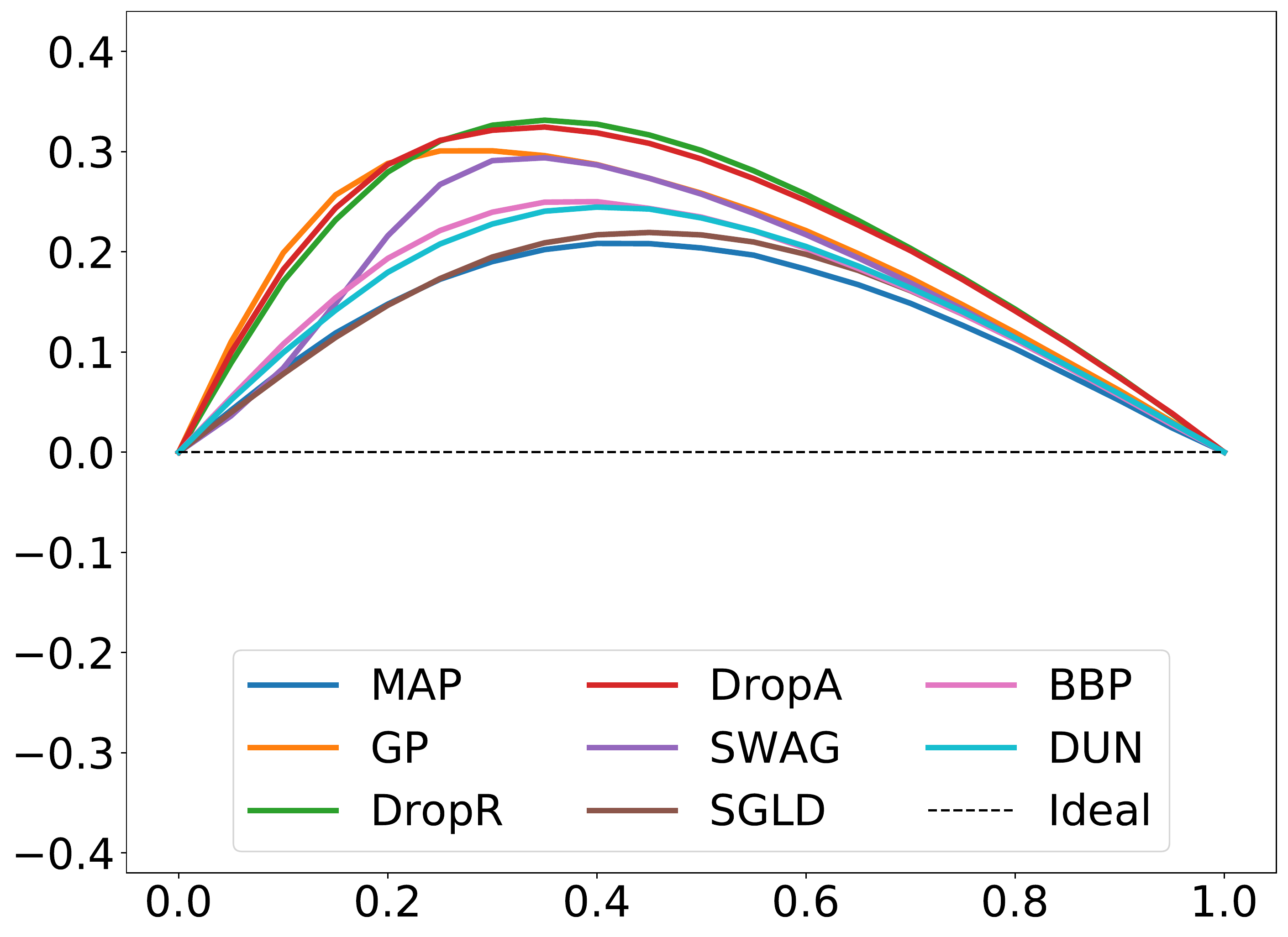}\\
\vspace{3mm}
\includegraphics[width=0.52\textwidth]{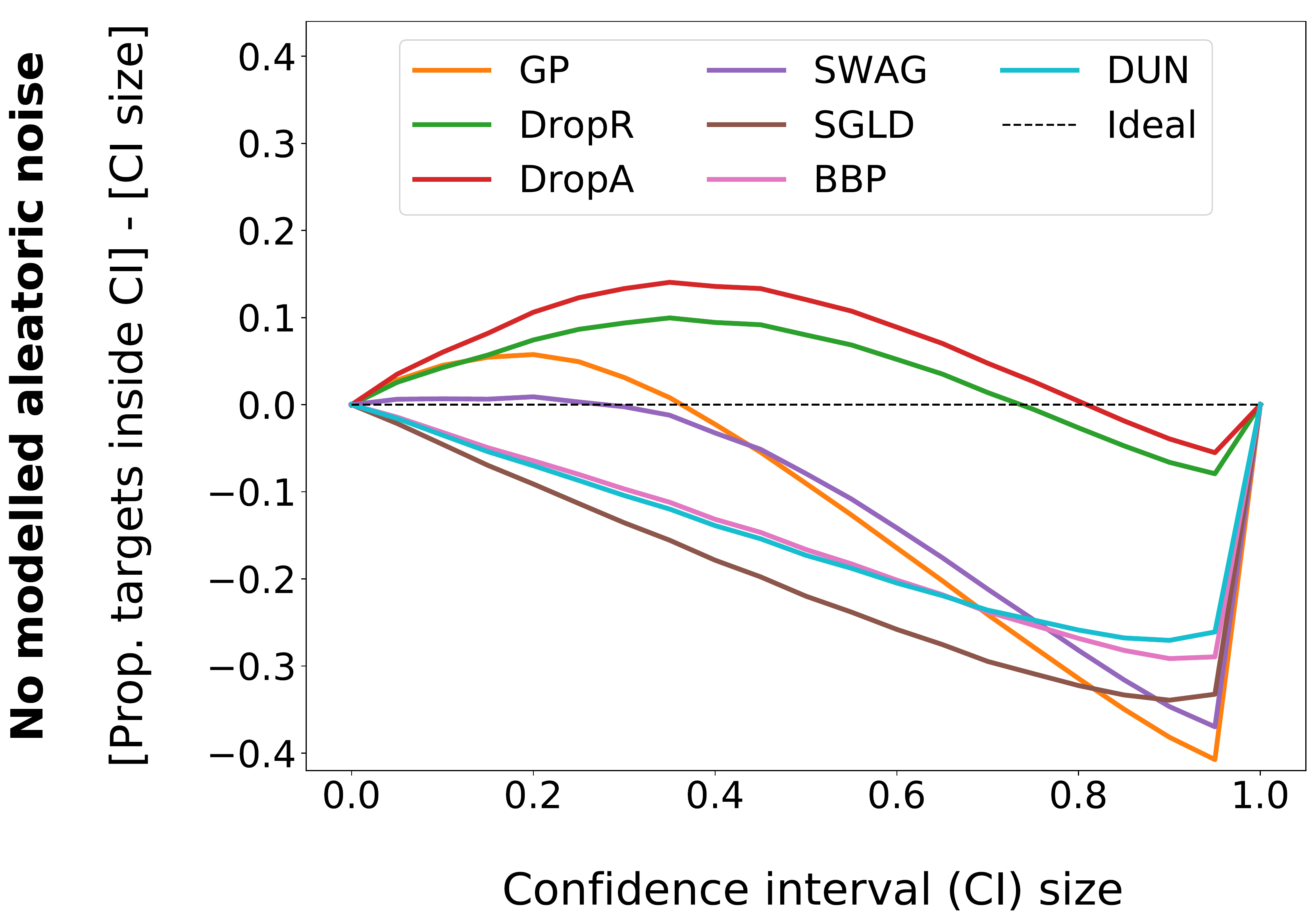}
\hfill
\includegraphics[width=0.4428\textwidth]{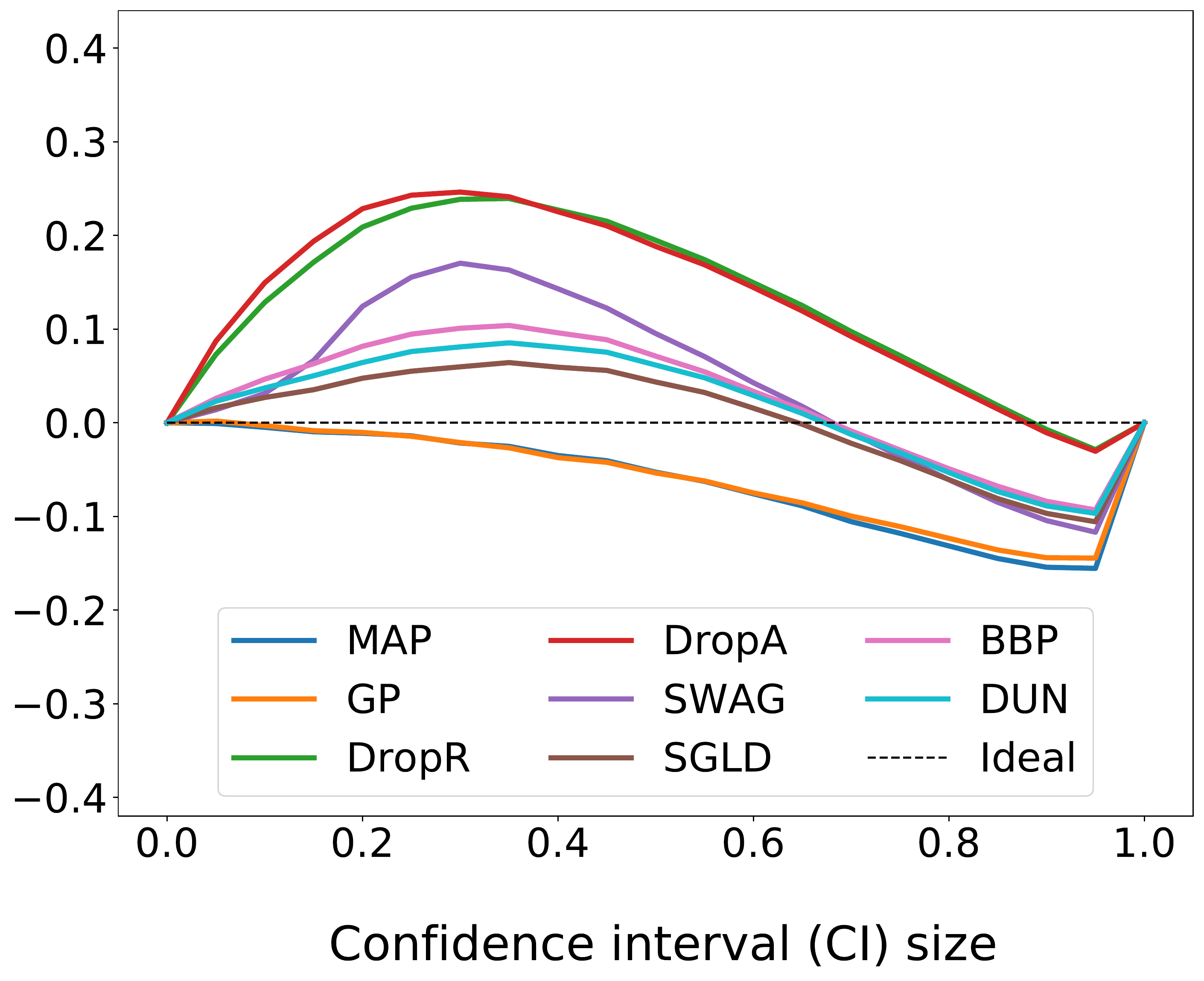}
\caption{Reliability diagrams for \textbf{single models} (left column) and \textbf{model ensembles} (right column). Each row (pair of diagrams) corresponds to a different likelihood function. Each line is the average of 5 runs.}
\label{fig:reliability_set}
\end{figure}

Figure \ref{fig:reliability_set} contains a complete set of reliability diagrams. Each row (pair of diagrams) corresponds to a different likelihood function. The first pair of diagrams are generated with Gaussian likelihoods; Gaussian noise parameters are learned when we train the D-MPNN. We observe pathological underconfidence across methods. The second pair of diagrams are generated with post-hoc $t$-distribution likelihoods and demonstrate improved calibration. The third pair of diagrams are generated without modelling aleatoric noise. In this case, the Bayesian predictive distribution is approximated as a single Gaussian rather than a mixture. We fit the single Gaussian to $S \times N$ predictive means, after drawing $S$ posterior samples from an ensemble of $N$ models. The third row demonstrates that overestimated aleatoric uncertainty drives the underconfidence in the first row. The elbow shape towards the end of the reliability lines in the third row points to the presence of outlying residuals.

\vspace{5mm}

\end{document}